\documentclass[pdflatex,sn-mathphys-num]{sn-jnl}% Math and Physical Sciences Numbered Reference Style

\usepackage{graphicx}%
\usepackage{subcaption} % For subfigures
\usepackage{multirow}%
\usepackage{amsmath,amssymb,amsfonts}%
\usepackage{amsthm}%
\usepackage{mathrsfs}%
\usepackage[title]{appendix}%
\usepackage{xcolor}%
\usepackage{textcomp}%
\usepackage{manyfoot}%
\usepackage{booktabs}%
\usepackage{algorithm}%
\usepackage{algorithmicx}%
\usepackage{algpseudocode}%
\usepackage{listings}%
\theoremstyle{thmstyleone}%
\theoremstyle{thmstyletwo}%

\theoremstyle{thmstylethree}%

\begin{document}

\title[Article Title]{Spatial-Wavelength Multiplexing Reliable Photonic Integrated General-Purpose Analog Computing System}


\author[1,2]{\fnm{Tao} \sur{Zhu}}
\equalcont{These authors contributed equally to this work.}

\author[1]{\fnm{Bowen} \sur{Zhu}}
\equalcont{These authors contributed equally to this work.}

\author[3]{\fnm{Shicheng} \sur{Zhang}}
\equalcont{These authors contributed equally to this work.}

\author[4]{\fnm{Keren} \sur{Li}}

\author[3]{\fnm{Xianchen} \sur{Wu}}

\author[1]{\fnm{Yazhi} \sur{Pi}}

\author[5]{\fnm{Jie} \sur{Yan}}

\author[5]{\fnm{Daigao} \sur{Chen}}

\author[3]{\fnm{Bingli} \sur{Guo}}

\author[1,5]{\fnm{Xi} \sur{Xiao}}

\author[1]{\fnm{Lei} \sur{Wang}}

\author*[1,2]{\fnm{Xiaochuan} \sur{Xu}}\email{xuxiaochuan@hit.edu.cn}

\author*[3]{\fnm{Xuwei} \sur{Xue}}\email{x.xue@bupt.edu.cn}

\author*[3]{\fnm{Shanguo} \sur{Huang}}\email{shghuang@bupt.edu.cn}

\author*[1]{\fnm{Zizheng} \sur{Cao}}\email{zcaopcl@foxmail.com}

\author*[1]{\fnm{Shaohua} \sur{Yu}}\email{yush@cae.cn}

\affil[1]{\orgname{Pengcheng Laboratory}, \orgaddress{\city{ShenZhen}, \postcode{518055}, \country{China}}}

\affil[2]{\orgdiv{School of Integrated Circuits}, \orgname{Harbin Institute of Technology (Shenzhen)}, \orgaddress{\city{Shenzhen}, \postcode{518055}, \country{China}}}

\affil[3]{\orgdiv{School of Electronic Engineering}, \orgname{Beijing University of Posts and Telecommunications}, \orgaddress{\city{Beijing}, \postcode{100876},\country{China}}}

\affil[4]{\orgdiv{College of Physics and Optoelectronic Engineering}, \orgname{Shenzhen University}, \orgaddress{\city{Shenzhen}, \postcode{518055}, \country{China}}}

\affil[5]{\orgname{National Optoelectronics Innovation Center}, \orgaddress{\city{Wuhan}, \postcode{430074}, \country{China}}}

\abstract{In the ``post-Moore era", the growing challenges in traditional computing have driven renewed interest in analog computing, leading to various proposals for the development of general-purpose analog computing (GPAC) systems.
In this work, we present a GPAC prototype featuring a silicon photonic chip designed for fully optical analog computation. This system leverages on-chip multi-channel architectures to enable parallel processing and utilizes wavelength-division multiplexing to significantly enhance computational capacity. In addition, we have developed an error-correction algorithm to monitor processing operations in real time, ensuring the reliability of computational results. Experimentally, we demonstrate the system's capability to solve ordinary differential equations and its applications in communications, microwave photonics, and image processing. The chip's energy efficiency is evaluated to reach up to 227 tera-operations per second per watt. Through this research, we provide a novel hardware framework and innovative directions for analog photonic computing.}




\maketitle

\section{Introduction}\label{sec1}

\hspace{2em}Analog computing was once widely used in the mid-20th century for solving differential equations and simulating dynamic systems. However, due to challenges like high noise sensitivity, unstable components, and limited scalability, it was overshadowed by digital systems, which, fueled by Moore’s law and advancements in semiconductor technology, have provided higher performance and lower costs. As a result, analog computing has long been sidelined in mainstream research.
With semiconductor processes approaching their physical limits and digital architectures facing energy and performance challenges, the “post-Moore’s law era” has reignited interest in alternative computing methods. Analog-like solutions, such as biological circuit, optical computing, and quantum computing, are gaining attention for their potential to bypass the von Neumann bottleneck and reduce digital logic power consumption~\cite{daniel2013synthetic, solli2015analog, ladd2010quantum}. However, many of these emerging options remain task-specific and not yet general-purpose. Additionally, quantum computing hardware and error-correction approaches are still maturing, limiting their widespread adoption.

In light of these challenges, researchers are exploring ways to combine the high parallelism and continuity inherent in analog computing with the pursuit of highly efficient and scalable computation. As early as 1941, Shannon introduced the General-Purpose Analog Computer (GPAC), proposing that ordinary differential equations (ODEs) could be mapped to analog circuitry using integrators, adders, and multipliers~\cite{shannon1941mathematical, gracca2003analog}. This concept has inspired recent advancements in neuromorphic computing and mixed-signal systems, though practical implementation is hindered by noise interference, component variations, and system reconfigurability issues, raising concerns about the reliability of analog computing.

To address these challenges, we propose a prototype GPAC based on photonic integrated circuits (PICs), incorporating specialized design and control strategies to improve noise resistance and mitigate non-ideal effects. Programmable PICs offer a highly parallel, energy-efficient, and fast solution for a broad range of computational tasks~\cite{bogaerts2020programmable, chen2018emergence}. PICs leverage photons as information carriers, enabling spatial routing and control of light through electrically tunable beam couplers connected by optical waveguides. These circuits support a variety of functions and have found applications in fields such as analog computing~\cite{ferrera2010chip, liu2017silicon}, differential equations solver\cite{tan2013high,yang2014all,yuan2025microcomb}, microwave photonics (MPW)~\cite{yao2022microwave,feng2024integrated,shu2022microcomb,wei2025programmable,perez2024general}, quantum information processing~\cite{wang2020integrated,zhu2024large}, matrix multiplexing~\cite{zhou2022photonic}, and photonic neural networks~\cite{shen2017deep,feldmann2021parallel,xu202111,bandyopadhyay2024single,bai2023microcomb}. The reconfigurability and high-speed processing of programmable PICs make them a versatile platform for GPAC systems.

Common programmable PICs typically consist of forward networks formed by optical waveguides interconnected via \(2\times 2\) Mach-Zehnder interferometer (MZI) optical switches in specific topologies, such as rectangular~\cite{du2024ultracompact, bandyopadhyay2024single} and triangular~\cite{zhang2021optical} configurations. These networks enable matrix operations and can be arranged into loop-based recurrent networks, forming two-dimensional (2D) grids, including rectangular~\cite{zhuang2015programmable}, triangular~\cite{perez2016reconfigurable}, and hexagonal~\cite{perez2020multipurpose, perez2017multipurpose} lattices. These grid structures allow for multi-directional light routing, enabling the realization of a fully programmable scattering matrix across all waveguide ports. Compared to forward networks, such grids support discrete-length delays (integer multiples), facilitating the creation of interferometers and resonant wavelength filters, like reconfigurable micro-ring resonators (MRRs)~\cite{liu2016fully}. This flexibility enhances the functional diversity and scalability of programmable PICs, making them ideal for advanced optical computing and signal processing applications.

However, fabrication errors inevitably arise during chip manufacturing due to limitations in processing precision. Silicon photonic chips are particularly sensitive to temperature variations, and the dense arrangement of components intended to minimize chip size exacerbates thermal crosstalk, impacting the optical phase of on-chip signals. Even small deviations in individual components can lead to significant errors in circuit computations. Consequently, developing error-correction capabilities for PICs has become a crucial research focus~\cite{xu2022self}, aiming to ensure robust and reliable performance in practical applications.

In this work, we demonstrate a silicon photonic chip that realizes a fully optical, general-purpose analog computing framework with real-time reliability control. By seamlessly integrating multi-channel architectures and wavelength-division multiplexing (WDM), our platform achieves a significantly enhanced computational throughput. Moreover, an FPGA-based feedback loop actively corrects thermal and component variations, addressing a long-standing barrier in analog computing. Through extensive experiments--ranging from solving ordinary differential equations and generating ultra-wide band (UWB) signals to demodulating WDM binary-phase-shift-keying (BPSK) data and high-speed images’ edge features detection--we show that photonic-based GPAC can be both robust and broadly applicable. These findings indicate a promising pathway toward energy-efficient, scalable analog computing solutions for next-generation communication, sensing, and AI acceleration. We believe this approach will benefit researchers in silicon photonics, neuromorphic computing, and beyond.

%Springer Nature does not impose a strict layout as standard however authors are advised to check the individual requirements for the journal they are planning to submit to as there may be journal-level preferences. When preparing your text please also be aware that some stylistic choices are not supported in full text XML (publication version), including coloured font. These will not be replicated in the typeset article if it is accepted. 

\section{Setup}\label{sec2}
\hspace{2em}The GPAC is a theoretical framework that models analog computation through a network of interconnected components. The primary components of the GPAC include adders, multipliers, integrators, and differentiators. These components are specifically designed to perform fundamental mathematical operations, which enables the simulation of a wide range of dynamic systems. The GPAC model is particularly adept at solving systems of ODEs, making it highly suitable for modeling physical phenomena in fields such as physics and engineering.

Shannon's work established that the functions computable by the GPAC are precisely those that are differentially algebraic~\cite{shannon1941mathematical, MacLennan1993ARO}. A function \( f: \mathbb{R} \to \mathbb{R}^k \) is generated by a GPAC if and only if all of its components satisfy a polynomial differential equation of the form:
\begin{equation}
    p\left( t, y, y', \dots, y^{(n)} \right) = 0,
\end{equation}
where \( p \) is a nonzero polynomial with real coefficients, and \( n \in \mathbb{N} \).

\begin{figure}[!h]
    \centering
    \includegraphics[width=1\textwidth]{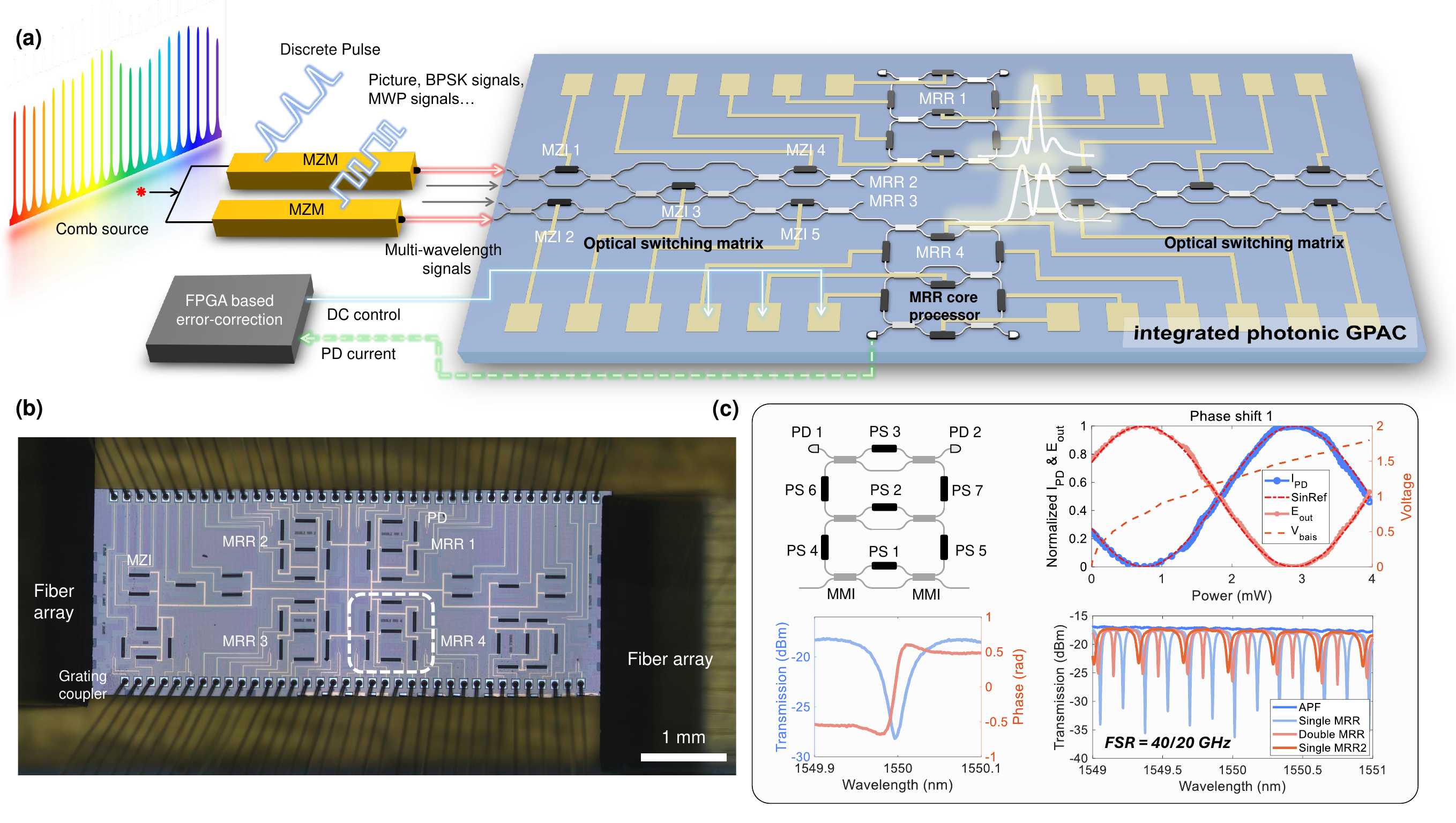}
    \caption{(a). Schematic diagram of spatial-wavelength multiplexing reliable photonic integrated GPAC system. (b). Microscopic image of the integrated silicon photonic circuit. (c). A detailed structural diagram of the MRR core processors. PS: phase shifter, MMI: multimode interference. The right panel of the figure presents the fundamental characterization results of the micro-ring resonator (MRR). By applying a voltage to phase shifter 1 (PS1), the optical power coupled into the micro-ring can be precisely tuned. Results demonstrate an inverse relationship between the photodetector (PD) current and the optical power at the Through port, with both parameters exhibiting a sinusoidal dependence on the phase shifter's power. The orange dashed line represents the voltage applied to PS1, where the square of the voltage is proportional to power. The lower-left image in the figure presents the simulated spectral response of the MRR. When the optical power output at the Through port reaches its minimum, a $\pi$-phase shift occurs, which is a critical feature enabling the use of MRRs for differentiation operations. The lower-right image displays the experimentally measured spectral response of the MRR. By adjusting the voltage applied to the phase shifter (PS), the MRR can be reconfigured into various operational modes, including an all-pass filter (APF), a single-ring structure, and a double-ring structure, with FSRs of 40 GHz and 20 GHz, respectively. }
    \label{fig:chip setup}
\end{figure}

Inspired by the principles of GPAC, we propose an innovative PIC that enables analog computation in the optical domain.
The proposed PIC features four parallel channels and reconfigurable MRRs, as shown in Fig.~\ref{fig:chip setup}(a). The architecture consists of three primary components, two optical switching matrices and a processing core unit in the middle. These three parts are connected in sequence, grating couplers are adopted at the input and output ports for the coupling of optical signals. 

The optical switching matrix is composed of five MZI optical switches through topological cascading, which enable the on-chip routing, switching, and merging of optical signals across four channels. 
Each MZI switch consists of two Mode Multiplexing Interferometers (MMI) and a thermal phase shifter. These switches can operate in three distinct states: bar state, cross state, and tunable coupler. By configuring the states of these optical switches, on-chip optical signal routing can be dynamically reconfigured to optimize performance and facilitate flexible analog computing operations. 

Each of the four channels features a core processing unit, which consists of a reconfigurable double ring structure, as shown in Fig.~\ref{fig:chip setup}(c). The MRRs regulate the coupling of optical signals via three tunable MZI switch structures, allowing for dynamic adjustment of optical properties. 
Specifically, the system supports two distinct free spectral ranges (FSRs), specifically \(40\) GHz and \(20\) GHz. This design enables 15 reconfigurable configuration states, including three types of all-pass filters.  
Fig.~\ref{fig:chip setup}(c) illustrates four spectral responses of these states. Additionally, four thermally tunable phase shifters on both sides of the micro ring can be adjusted to shift the spectral response of the micro ring red or blue on the spectrum. Meanwhile, there is an integrated Ge photodetector (PD) at each of the two drop ports of the micro ring, respectively. These PDs are utilized to detect the intensity of optical signals, enabling the state of the micro rings to be locked and controlled through feedback. 

The core processing unit can be configured as a time-domain optical differentiator, enabling first-order differentiation of input optical signals~\cite{liu2008compact}. The frequency-domain transfer function is given as $T_{\omega} = j(\omega – \omega_0)$, where $\omega_0$ represents the optical carrier frequency. 
The four-channel optical carrier signals, which have been modulated to carry data information, enter the processing unit from the left input ports.
The input optical signals can be represented as a \(4 \times 1\) vector. 
Furthermore, each channel of the mixed optical signals can simultaneously contain wavelengths ranging from $\lambda_1$ to $\lambda_4$. After propagating through the \( 4 \times 4 \) switching matrix, matrix operations are performed across all wavelengths simultaneously, achieving the space-frequency interleaved computation paradigm. 

\section{Reliable general purpose analog computing}\label{sec3}

\hspace{2em}Shannon's GPAC produces solutions of ODEs as its output and serves as a distinguishing feature. In this section, we will implement ODE solving with our chip. Before that, we introduce the basic operations that serve as fundamental components of the GPAC paradigm.

\begin{figure}[!h]
    \centering
    \includegraphics[width=0.8\textwidth]{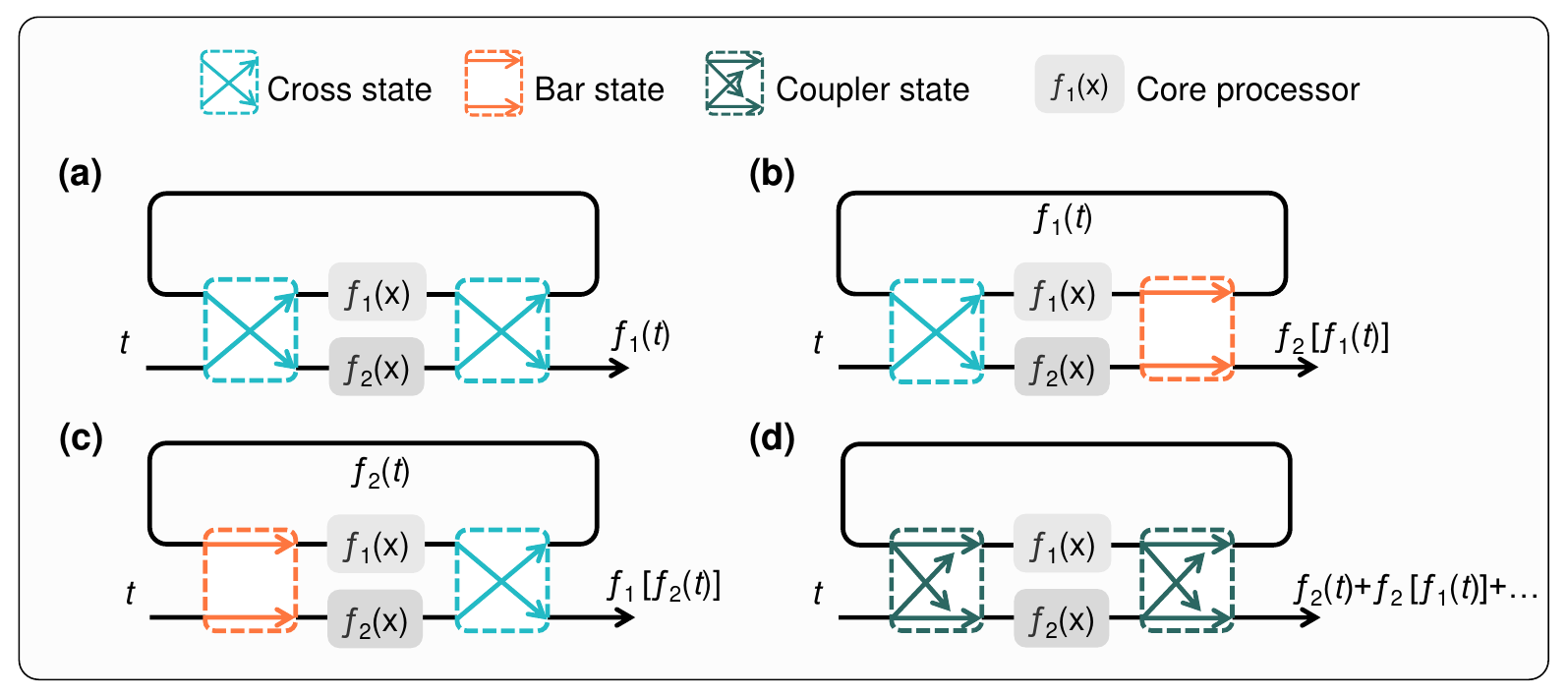}
    \caption{Schematic diagram of spatial multiplexing configurations.}
    \label{fig:spatial multiplexing}
\end{figure}

The proposed chip can be configured to achieve complex computing functions, which can be equated to spatial multiplexing, are illustrated in Fig.~\ref{fig:spatial multiplexing}. Taking the two-channel case as an example, since the MZI optical switches in optical switching matrices can operate in three different working states, the on-chip exchange and coupling of optical signals can be realized by utilizing these MZI optical switches. Suppose that the processing of the input optical signal $x(t)$ by the core processing units of the two channels is denoted as $f_{1}(x)$ and $f_{2}(x)$, respectively. The simplest case is to perform the processing on the input optical signal only once, as shown in Fig.~\ref{fig:spatial multiplexing}(a). At this case, the chip enables multi-channel parallel computing, which enhances the computational efficiency of the system. Meanwhile, the higher the extinction ratio of the MZI optical switches is, the smaller the crosstalk between channels will be, and the less impact it will have on the computational results. 

The second case is that the results of the computation are transmitted back onto the chip through a loop back to continue the second-step computation, as shown in Fig.~\ref{fig:spatial multiplexing}(b) and~\ref{fig:spatial multiplexing}(c). By adjusting the MZI optical switches before and after the core processing unit, the sequence of the computation process can be manipulated. The output results are precisely those of the two-step computation. Utilizing this structure enables the implementation of single-chip operations such as second-order or even higher-order differential computations. 

The last case involves more complex situations, as shown in Fig.~\ref{fig:spatial multiplexing}(d). By configuring the MZI optical switches into the coupling state (tunable coupler, TC), the splitting and combining of on-chip optical signals can be achieved, thus the output result is the sum of two computational results. In fact, the structure depicted in Fig.~\ref{fig:spatial multiplexing}(d) forms a resonant system similar to a MRR. When the input optical signal is a pulsed signal, the output of the system corresponds to the steady-state solution of the input optical signal within the resonant structure. This configuration enables the implementation of calculations such as Taylor series expansion, as well as the solution of ODEs with constant coefficients. 

\begin{figure}[!h]
    \centering
    \includegraphics[width=1\textwidth]{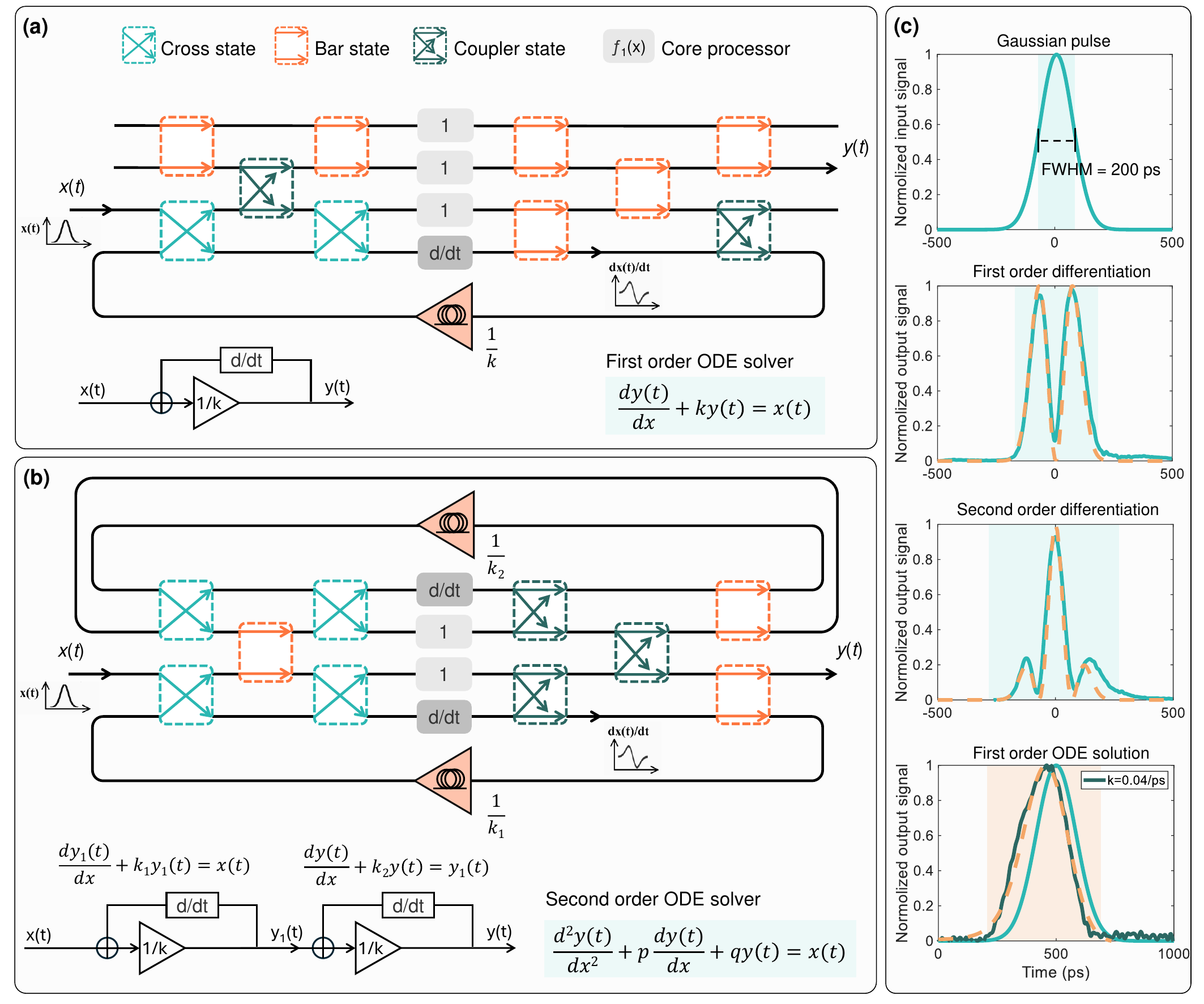}
    \caption{Schematic diagram of GPAC configured as ODEs solver utilizing real time differential operations. (a). Configuration of a first-order ODE solver. (b). Configuration of a second-order ODE solver. (c). waveform results of first-order and second-order differentiation applied to a Gaussian pulse with a FWHM of 200 ps, and the first-order ODE result with a Gaussian pulse as input $x(t)$. }
    \label{fig:ODE}
\end{figure}

% As a GPAC chip, its distinguishing feature lies in the photonic domain implementation of fully optical ODEs solving. ODEs are extensively employed to model physical systems, such as mechanical motion dynamics and electrical circuit modeling. 
The proposed chip leverages core processing units to perform differential operations, with configurable MZI optical switch and loop back amplification interconnects enabling programmable ODE solutions, as illustrated in Fig.~\ref{fig:ODE}(a). The lower-left schematic illustrates the solver system architecture. The system propagates the input pulse signal $x(t)$ into a feedback loop, where iterative signal processing ensures convergence to a stable state, and the output signal $y(t)$ is the solution to the equation of $\frac{\mathrm{d} y}{\mathrm{d} x} + ky(t) = x(t)$. The $d/dt$ block in the diagram represents the differentiation operation performed by a reconfigurable MRR. The $1/k$ coefficient is adjusted via an erbium-doped fiber amplifier (EDFA), while optical signal combining is achieved through on-chip MZI optical switches.  Upon achieving loop stabilization, the solution $y(t)$ to the first-order ODE is obtained. The detailed mathematical derivation of this process, including the formulation of the feedback dynamics and stability criteria, is provided in the Methods. This approach leverages the inherent properties of the feedback architecture to enable real-time analog computation of ODE solutions, offering a robust alternative to traditional numerical methods. Thus, achieves high-speed, energy-efficient computation by harnessing photonic signal propagation for analog differentiation, thereby bypassing the quantization limitations inherent to digital processing frameworks. 

The GPAC chip can also be configured for solving second-order ODEs, as illustrated in Fig.~\ref{fig:ODE}(b). This functionality is achieved by cascading two first-order ODE solvers, where the output of the first solver serves as the input to the second. Leveraging the chip's four-channel architecture, this configuration requires only three feedback loops to implement. Extending this principle, higher-order ODEs can be solved by cascading multiple chips, enabling scalable computational capabilities. However, this approach is significantly impacted by photonic signal noise, as the amplifiers within each feedback loop amplify noise cumulatively. The noise effect becomes increasingly pronounced with additional cascading stages. To mitigate this issue, the design of on-chip feedback loops presents an effective solution, as it minimizes noise propagation while maintaining signal integrity across multiple processing stages. Fig.~\ref{fig:ODE}(c) demonstrates the waveform results of first-order and second-order differentiation applied to a Gaussian pulse with a full width at half maximum (FWHM) of 200 ps. Additionally, the figure presents the solution to a first-order ODE. 

\begin{figure}[!h]
    \centering
    \includegraphics[width=1\textwidth]{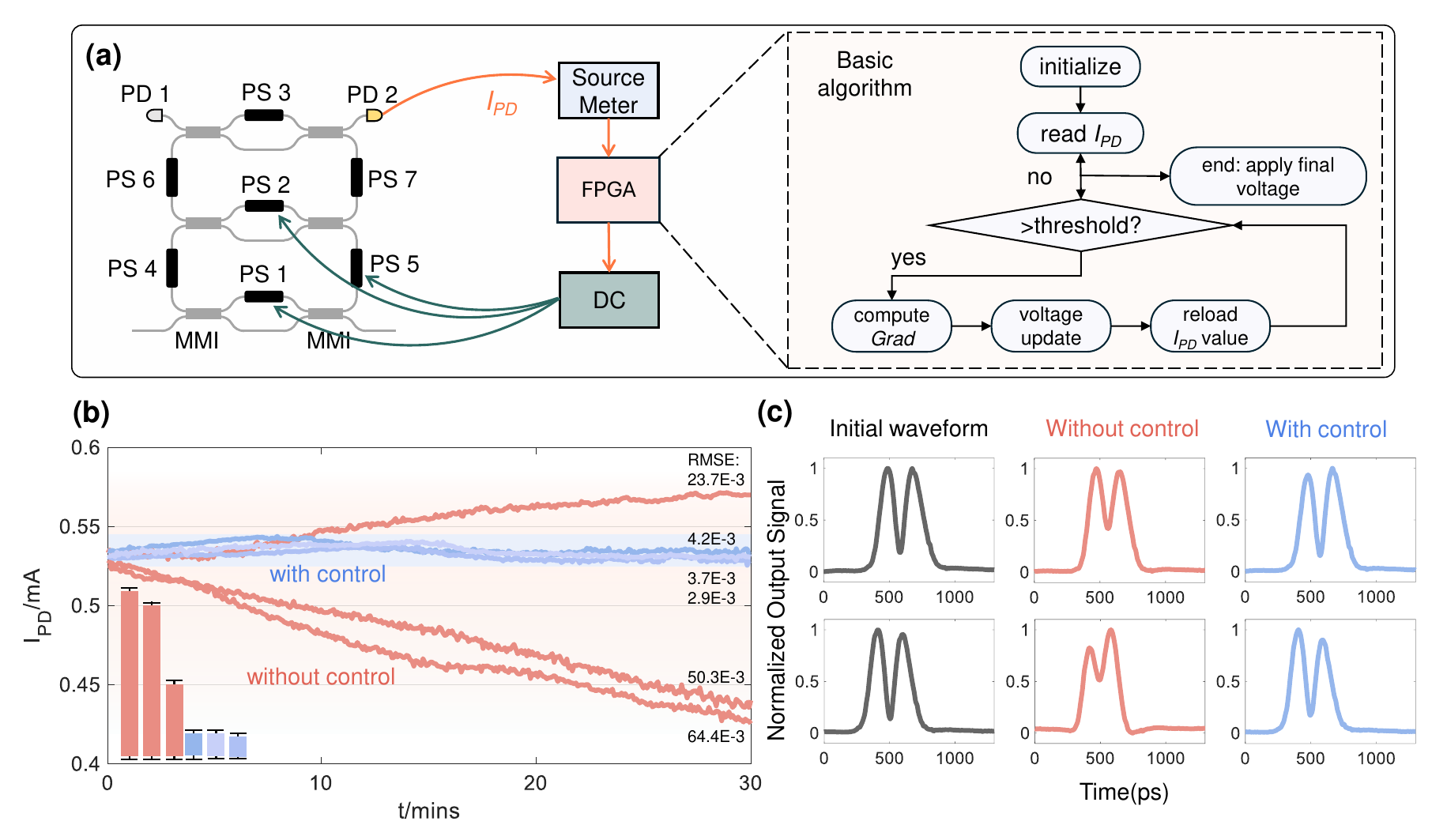}
    \caption{FPGA based GPAC error-correction.(a). Schematic diagram of FPGA based real-time error-correction system. (b). PD current values recorded for 30 minutes with/without error-correction algorithm. (c). Oscilloscope waveforms of a first-order differentiation result of a Gaussian pulse with/without error-correction algorithm after 30 minutes. }
    \label{fig:control}
\end{figure}

Enhancing the reliability of computational results is a critical challenge in analog computing, with the credibility of the computation process being one of its core objectives. In our GPAC chip, the core processing unit incorporates an on-chip PD. By continuously monitoring the PD current in real time, the system achieves error-correction capabilities during processing, thereby ensuring the accuracy of computational outcomes. A detailed description of this error - correction algorithm can be found in Method. We evaluated the error-correction characteristics of the system, as illustrated in Fig.~\ref{fig:control}(b). The PD current serves as an indicator of the operational state of the reconfigurable MRR within the core processing unit. As we all know, silicon photonic chips are susceptible to temperature variations, which induce drift in the MRR's operational state. As shown in the plotted curves, we recorded three sets of PD current values with the error-correction algorithm enabled and three sets without the algorithm, while the micro-ring was maintained in a differentiation-computation state for 30 minutes. When the error-correction algorithm was enabled, the fluctuations in the three sets of current values remained within a range of 0.02 mA. In contrast, the other three sets of current values exhibited continuous drift without correction, reflecting the overall instability of the microwave photonic experimental link. This behavior is most obviously observed in the oscilloscope waveforms, as shown in Fig.~\ref{fig:control}(c). At the beginning of the evaluation, the computational output waveform corresponded to the first-order differentiation result of a Gaussian pulse. However, after 30 minutes without the error-correction algorithm, the output waveform exhibited significant deviations, indicating that the reconfigurable MRR was no longer operating in a first-order differentiation mode but potentially in a fractional-order differentiation state. This deviation rendered the computational results unreliable, significantly compromising the accuracy of the analog processing outcomes. In contrast, with the error-correction algorithm active, the output waveform remained stable. This approach not only mitigates the inherent noise and variability in analog systems but also establishes a robust framework for reliable and verifiable analog computation. 

\section{Experiments and results}\label{sec4}

\hspace{2em}We first experimentally demonstrated the generation of ultra-wide band (UWB) signals based on differential operations, which provides a novel approach for the generation of UWB signals in the optical domain. UWB signals possess an extremely wide bandwidth and an extremely narrow signal pulse width. Under the same signal-to-noise ratio, they can offer a higher channel capacity, which makes them applicable to the rapid transmission of large amounts of data within short distances. Moreover, UWB signals can also conduct high-resolution detection on targets at close ranges~\cite{pan2010uwb,pan2009photonic,li2009millimeter}. 

\begin{figure}[!h]
    \centering
    \includegraphics[width=1\textwidth]{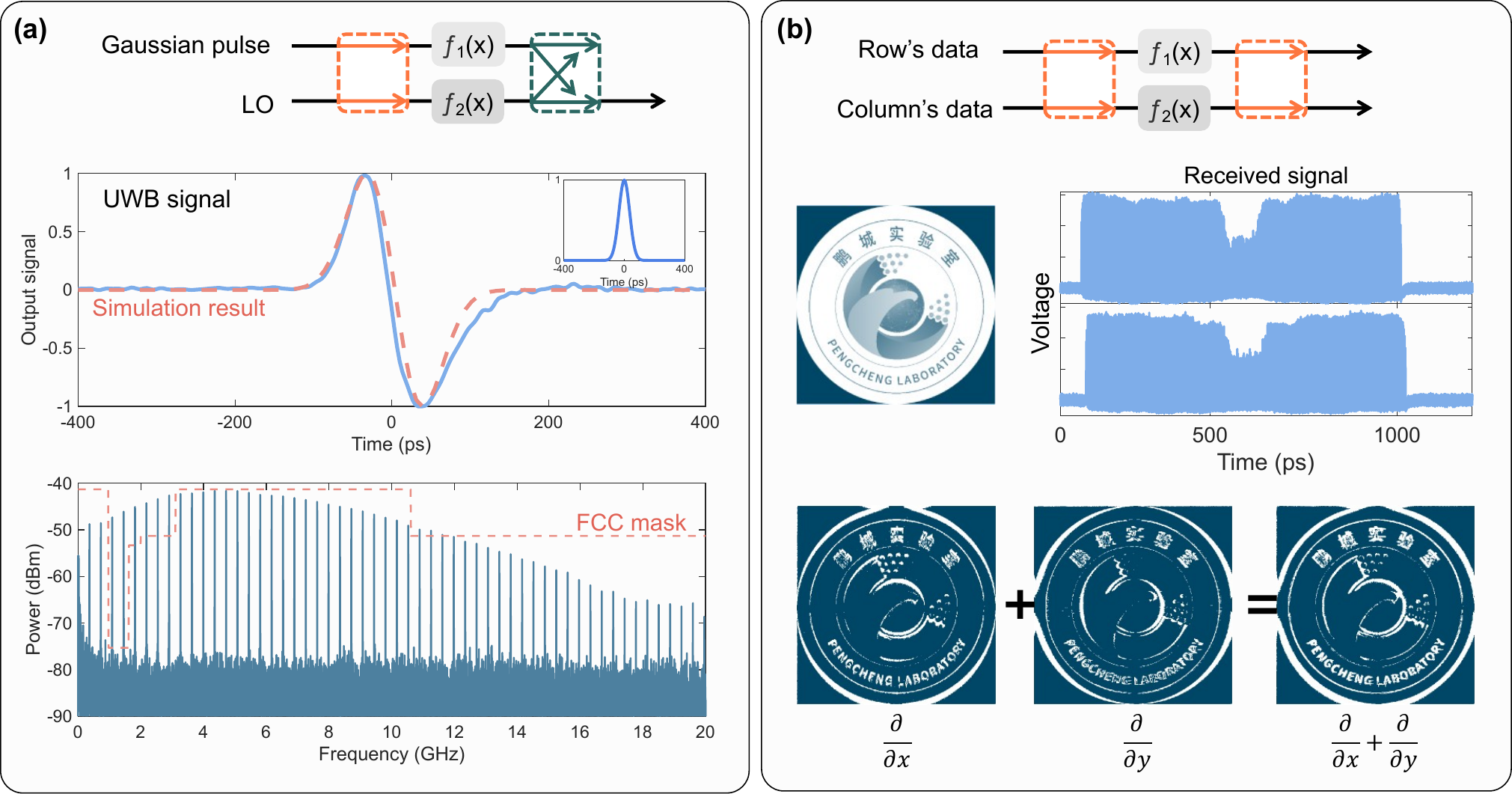}
    \caption{UWB signal generation and high-speed images’ edge features detection utilizing spatial multiplexing. (a). The upper portion illustrates the on-chip configuration for UWB signal generation, the middle illustrates the time-domain waveform of the output UWB signal, insert showing the input Gaussian pulse waveform. The lower portion displays the corresponding frequency spectrum of the synthesized UWB signal, with FCC mask.  (b). The upper portion illustrates the on-chip configuration for image edge detection, the middle shows the original picture and received signals, the lower portion displayed the results.}
    \label{fig:UWB&edge detection}
\end{figure}

UWB signals can be accomplished by utilizing on-chip combination approach, as shown in Fig.~\ref{fig:UWB&edge detection}(a). A Gaussian pulse with a FWHM = 100 ps is introduced into one channel on the chip, while an optical signal of the same wavelength is fed into the other channel as the local oscillator signal. At this point, the core processing unit $f_{1}(x)$ performs a first-order differential operation, and the other processing unit $f_{2}(x)$ does not perform any processing, which is equivalent to $f_{2}(x) = 1$. 

Another application of GPAC is images edge detection, as shown in Fig.~\ref{fig:UWB&edge detection}(b). Initially, the grayscale information of the image is extracted.  The grayscale image is subsequently one-dimensionalized in a row-by-row or column-by-column manner to generate continuous one-dimensional image information. By performing differential processing on the image signal, those parts with drastic changes in grayscale values can be calculated, thus achieving edge detection. Thereafter, the computational results are reconstructed into a two-dimensional image, which serves as the outcome of the image edge detection. However, this method suffer an issue that When the image is one-dimensionalized row-by-row, the computational process cannot effectively detect the changes in grayscale values of the original image in the column direction. A promising solution is to one-dimensionalize the image both row-by-row and column-by-column, and simultaneously perform differential calculations on the information in both directions. Finally, by combining the two results, the obtained detection outcome exhibits a better performance. 

A notable advantage of the PIC chip is its capability for parallel processing in the frequency domain, which allows it to significantly enhance processing efficiency in a highly compact, energy-efficient, and cost-effective manner. To verify this, we demonstrated the chip's ability to simultaneously demodulate WDM BPSK signals at the receiver end in high-speed optical communication applications.

In optical fiber communication, BPSK modulation offers the same spectral efficiency as On-Off Keying(OOK) but exhibits better tolerance to nonlinear distortions and higher receiver sensitivity. However, since the carrier is suppressed, demodulating BPSK signals requires complex coherent receivers or precisely controlled delay-line interferometers (DLIs). By configuring MRRs as a temporal differentiator to replace the DLIs, differential detection of BPSK signals can be achieved~\cite{zhang2007microring,padmaraju2012error,xu2009optical}. In ~\cite{li201410}, the demodulation of a 10 Gbps BPSK signal using a MRR was demonstrated for the first time. In our experiment, we showcased the MRR demodulation of single-wavelength 32 Gbps BPSK signal and \(5\times 25\) Gbps WDM BPSK signals, validating the chip's application in optical communication and its capability for parallel processing in the frequency domain.

\begin{figure}[!h]
    \centering
    \includegraphics[width=0.8\textwidth]{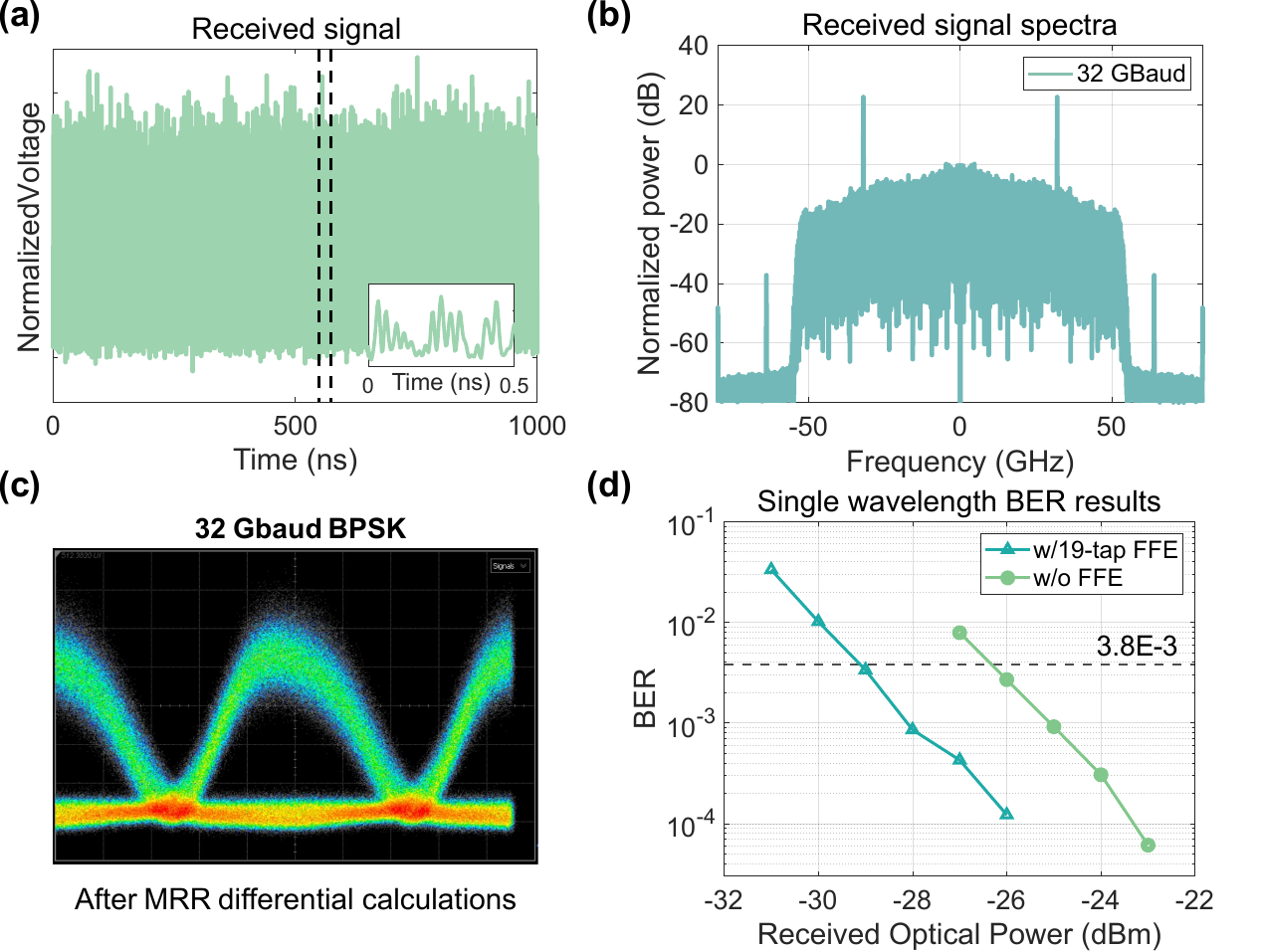}
    \caption{Demodulation of single-wavelength 32Gbaud BPSK signal. (a)(b). Received demodulated BPSK signal and its spectra diagram.  (c). Eye diagrams of the 32 GBaud BPSK signal after MRR differential calculation. (d). The BER result with/without a simple 19-tap linear equalizer. }
    \label{fig:bpsk}
\end{figure}

Before conducting the WDM tests, we first tested the demodulation of single-wavelength 32 Gbaud BPSK signal to verify feasibility. In this case, only CW light was input into the second MZM. Fig.~\ref{fig:bpsk}(a) and (b) show the eye diagrams of the signals before and after MRR processing. 

Fig.~\ref{fig:WDMbpsk}(a) illustrates the experimental setup. Here, a processing unit of the PIC chip was activated, and its reconfigurable double ring was configured as a single MRR with an FSR of 40 GHz, enabling it to handle BPSK signals with sufficiently large bandwidths. For the WDM case, the first single-drive MZM was driven by a 40 GHz clock to generate an optical frequency comb with intervals corresponding to the FSR of MRR. After amplification, five lines of the frequency comb served as carriers and were input into the second MZM for simultaneous modulation, generating five 25 Gbaud WDM BPSK signals. At the receiver, an optical attenuator (ATT) was used to adjust the received power. After processing the WDM BPSK signals through the MRR, a wavelength selective switch (WSS) selected the target wavelength signal, which was then converted into an electrical signal by a PD and captured by an oscilloscope. 

Fig.~\ref{fig:bpsk}(b) shows the relationship between the bit error rate (BER) and received optical power. After applying a simple 19-tap linear equalizer, a receiver sensitivity of -29 dBm was achieved at the 7$\%$ HD-FEC threshold. In the WDM experiment, Fig.~\ref{fig:WDMbpsk}(b) display the signal spectra before and after MRR processing, with a resolution of 0.1 nm. We measured the BER results for each signal, as shown in Fig.~\ref{fig:WDMbpsk}(b). In the WDM case, the receiver sensitivity was approximately -24 dBm.

\begin{figure}[!h]
    \centering
    \includegraphics[width=1\textwidth]{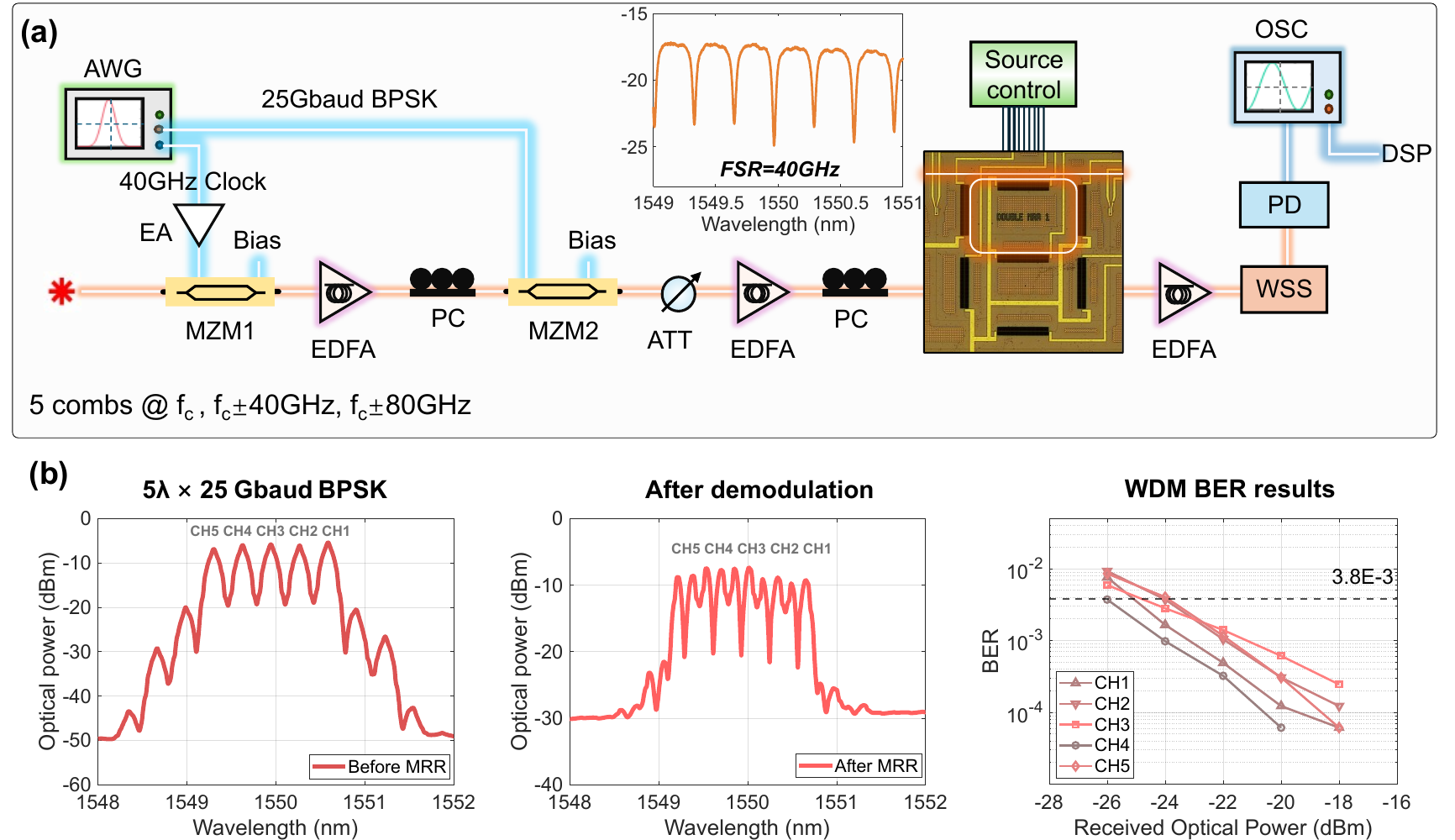}
    \caption{Demodulation of \(5\times 25\) Gbps WDM BPSK signals. (a). Experimental setup, insert is the spectra response of the core processor MRR. EA: electrical amplifiers; PC: polarization controller; ATT: attenuator; DUT: device under test; WSS: wavelength selective switch; PD: photodetector; AWG: arbitrary waveform generator; OSC: oscilloscope. (b). Signal spectra before and after MRR processing and the BER results. }
    \label{fig:WDMbpsk}
\end{figure}

\section{Discussion}\label{sec5}

\hspace{2em}A comparative analysis of state-of-the-art analog computing chips is summarized in Table \ref{tab1}. In recent years, PIC based on heterogeneous integration platforms, such as thin-film lithium niobate (TFLN), have been extensively investigated. These platforms leverage the intrinsic material advantages (e.g., high electro-optic coefficients and low optical losses) to achieve significantly enhanced operational bandwidths. However, the critical challenge of computational reliability—particularly analog computing—remains unresolved in existing systems. Our proposed architecture using a FPGA based real-time error-correction algorithm, ensuring environmental robustness and long-term stability. Moreover, the multi-channel parallel processing design of our chip enhances computational efficiency by enabling simultaneous wavelength- and space-division multiplexing, thereby maintaining signal fidelity and anti-interference capability. In addition, the ability to cascading architecture allows flexible reconfiguration for diverse applications, including microwave photonic signal processing, optical image edge detection, and coherent communication systems.

\begin{sidewaystable}
\caption{Performance comparison of integrated analog computing circuits}\label{tab1}
\begin{tabular*}{\textheight}{@{\extracolsep\fill}p{1.5cm}p{1.5cm}p{1.5cm}p{1.5cm}p{1.5cm}p{1.5cm}p{1.5cm}p{2cm}p{1.5cm}p{2cm}}
\toprule%
& Material platform & System bandwidth & Basic processor & Shortest processable input pulse & Analog signal sampling rate & Energy efficiency & Robustness of computing & Capability to cascade & Multi-function\\
\midrule
2016\cite{liu2016fully} & Si & 55 GHz & Tunable MRR & 33 ps & N/A & N/A & N/A & N/A & N/A \\
2018\cite{zhang2018fully} & Si & 40 GHz & reconfigurable grating & 290 ps & N/A & N/A & N/A & Yes & N/A \\
2020\cite{zhang2020photonic} & Si & 40 GHz & FPDA & 85 ps & N/A & N/A & N/A & Yes & Wavelength multiplexing \\
2024\cite{lin2024120} & TFLN & 60 GHz & MZI & N/A & N/A & 153 TOPS/W & N/A & N/A & TDM \\
2024\cite{feng2024integrated} & TFLN & $>$67 GHz & MRR/MZI & 9.6 ps & 256 GSa/s & N/A & N/A & N/A & N/A \\
2025\cite{qian2025analog} & Si & $<$20 GHz & MZM+TDL & 25 ps & 65GSa/s & N/A & N/A & N/A & Parallel processing \\
\midrule
This work & Si & 40 GHz & Tunable MRR & 31 ps & 128 GSa/s & 227 TOPS/W & FPGA based real-time correction & Higher-order ODE solver & Spatial-wavelength multiplexing\\
\botrule
\end{tabular*}
\footnotetext{N/A=Information not available or not applicable.}
\footnotetext{TDL=tunable delay line, TDM=Time-division multiplexing. }
\end{sidewaystable}

\section{Conclusion}\label{sec6}

\hspace{2em}In this work, we propose a novel concept of implementing a GPAC using PIC and demonstrate a silicon photonic chip design aligned with this paradigm. The chip is capable of performing fully optical analog computations, while a FPGA-based error-correction algorithm ensures the reliability of the processing operations. Long-duration calibration procedures were validated through extensive testing. The core processing units, based on MRR structures, are reconfigured to adapt to various operational states, enabling fully optical analog computation with frequency-domain parallel processing. We demonstrate a case of solving an ODE based on analog differential computing, showcasing the completeness and functionality of the system as a GPAC implementation. Furthermore, we present three case studies for the chip's capabilities in optical communications and microwave photonics, including all optical UWB signal generation, high-speed images’ edge features detection utilizing spatial multiplexing, and demodulation of \(5\times 25\) Gbps WDM BPSK signals. By exploiting the periodic frequency-domain response of the core processing units, the chip enables parallel processing of multi-wavelength signals, facilitating wavelength-division multiplexing and significantly enhancing computational throughput and energy efficiency. Based on estimations, the chip achieves a computational efficiency of approximately over 200 TOPS/W, offering a new architectural framework and innovative directions for photonic computing. We believe that this study contributes to the ongoing exploration of general-purpose analog computing in the post-Moore’s law era and provides a foundation for future scalable GPAC-like architectures.

\section{Methods}\label{sec6}

\textbf{Device fabrication}

The integrated photonic chip was fabricated on 8” wafer using a commercial 220 nm silicon-on-insulator (SOI) platform, with a $3$ $\mu$m thick BOX. The chip has a total size of $5.9 \times 2.3$ mm. Silicon strip waveguide features a width of $500$ nm is patterned on a SiN hardmask using ArF lithography scanner and SiN dry etch, and its propagation loss is lower than $1.5$ dB/cm at wavelength of $1550$ nm. The length of each ring resonator in the core processor is $1.757$ mm, each MMI coupler in the MZI structure has $0.3$ dB insertion loss. The vertical grating coupler has a $10^{\circ}$ incident angle to maximum the coupling efficiency, and the insertion loss of the grating coupler is $<4$ dB at C-band. On chip photodetectors are fabricated using a $500$ nm thickness Ge with responsivity $>0.8$ A/W, the measured dark current of high-speed PD is about $20$ $\mu$A, and its O/E $3$ dB bandwidth is larger than $65$ GHz. To ease the experiment, two fiber arrays (FA) are utilized to effectuate the coupling of the signal light from the fiber to the chip. Additionally, on-chip DC pad and the DC connector on the printed circuit board (PCB) are interconnected through wire bonding. \\

\noindent\textbf{Principle of differential based ODE solver}

More details are presented here for principle of differential-based ODE solver. An optical temporal differentiator computes the first derivative of the optical field in the time domain, exhibits a transfer function of the form $T_{\omega} = j(\omega - \omega_0)$ in the frequency domain, where $\omega_0$ represents the optical carrier frequency. This transfer function indicates that the transmission of the differentiator is linearly proportional to the frequency detuning from the central frequency, while the phase response undergoes an exact $\pi$-phase shift across the central frequency. The core processing unit in Fig.~\ref{fig:chip setup}(c) can be reconfigured by adjusting the phase shifter (PS) to form a MRR coupled to a single straight waveguide. This reconfigurable ability enables precise control over the optical signal processing capabilities of the unit, making it a versatile component for applications requiring tailored spectral and phase responses. 
The transfer function of this configuration can be expressed as:
\begin{equation}
T(\omega)=\frac{S_0}{S_i}=\frac{j(\omega-\omega_0)+\frac{1}{\tau_i}-\frac{1}{\tau_e}}{j(\omega-\omega_0)+\frac{1}{\tau_i}+\frac{1}{\tau_e}}
\end{equation}
where $\omega_0$ represents the resonance frequency, $1/\tau_i$ represents the power decay rate due to the intrinsic loss, $1/\tau_e$ represents the power coupling to the waveguide. Thus the reciprocal of photon lifetime can be expressed as $1/\tau = 1/\tau_i +1/\tau_e$. Under the condition that the frequency detuning is significantly smaller than the 3-dB bandwidth of the resonator, the expression can be approximated as follows: 
\begin{equation}
T(\omega)=j\tau(\omega-\omega_0)+\frac{\frac{1}{\tau_i}-\frac{1}{\tau_e}}{\frac{1}{\tau_i}+\frac{1}{\tau_e}}
\end{equation}
Under this condition, the MRR coupled to a single straight waveguide can be modeled as a temporal differentiator with certain gain plus a constant-output. In particular, when the microring resonator works in the critical coupling region $\tau_i=\tau_e$, we obtain $T_{\omega} = j\tau (\omega - \omega_0)$, which is a typical function for a first-order temporal differentiator. 

The ODE solver configuration based on temporal differentiator presented in Fig.~\ref{fig:ODE}(a) can be achieved by spatial multiplexing utilizing MZI optical switches. The constant-coefficient first-order linear ordinary differential equation can be expressed as: 
\begin{equation}
\frac{\mathrm{d} y(t)}{\mathrm{d} t} +x(t)=ky(t),
\end{equation}
where $x(t)$ represents the input signal, $y(t)$ is the equation solution (output signal) and $k$ represents a positive constant of an arbitrary value. The equation can be numerically transformed to spectral domain using Fourier transformation:
\begin{equation}
Y(\omega )=\frac{1}{k-j\omega } · X(\omega )
\end{equation}

In time domain, the equation can be expressed as:
\begin{equation}
y(t)=u(-t) · e^{kt} \otimes x(t)
\end{equation}

The equation also can be expanded using Taylor expansion as follow: 
\begin{equation}
Y(\omega )=\frac{1}{k-j\omega } · X(\omega )=\frac{X(\omega )}{k} · \sum_{n=0}^{\infty } \left ( \frac{j\omega }{k}\right )^{n}
\end{equation}

Therefore, the time-domain representation of the equation can be derived by applying the inverse Fourier transform:
\begin{equation}
y(t)=\frac{1}{k} \sum_{n=0}^{\infty } (\frac{1}{k} )^{n}\frac{d^{n}x(t)}{dt^{n}}
\end{equation}

It can be seen from the formula that the equation converges only when $|j\omega /k|<1$. In fact, during the experimental process, there is a threshold for $k$. When this threshold is exceeded, a solution to the ODE cannot be obtained.

The operational principle of the ODE solver depicted in the Fig.~\ref{fig:ODE}(a) is as follows: In the first cycle, only the input signal $x(t)$  propagates within the loop. Consequently, $y(t)$  should be expressed as:
\begin{equation}
y_1(t)=\frac{1}{k} x(t)
\end{equation}

for the second circle, $y(t)$  should be expressed as:
\begin{equation}
y_2(t)=\frac{1}{k} (x(t)+\frac{dy_1(t)}{dt})=\frac{1}{k}(x(t)+ \frac{1}{k}\frac{dx(t)}{dt})
\end{equation}

When the loop eventually reaches a stable state, the output expression can be written as: 
\begin{equation}
y_n(t)=\frac{1}{k} (x(t)+\frac{dy_{n-1}(t)}{dt})=\frac{1}{k}(x(t)+\frac{1}{k}\frac{dx(t)}{dt}+\dots +(\frac{1}{k})^{n-1}\frac{d^{n-1}x(t)}{dt^{n-1}})
\end{equation}
The derived expression is consistent with the Taylor series expansion presented earlier, representing the solution to the ODE under the input $x(t)$. This alignment confirms the mathematical validity of the approximation and its applicability to modeling the system's response to time-varying inputs. For higher - order ODE equations, it only requires taking the solution $y_1(t)$ of the first ODE as the input $x_2(t)$ of the second solver. The principle of each solver is the same. \\

\noindent\textbf{Principle of BPSK signal receiving using MRR}

More details are presented here for principle of BPSK signal receiving using MRR. Figure~\ref{fig:WDMbpsk} presents the time-domain waveform of an original BPSK signal, along with the waveform after MRR-based differential transformation and PD detection from left to right. The corresponding spectral are shown below the waveforms. It can be observed that after differentiation, the BPSK waveform primarily consists of impulse-like functions at symbol transitions, with amplitudes depending on the direction of bit changes. This type of signal is also referred to as an alternate-mark inversion (AMI) signal. Since differentiation can be regarded as a notch filter, the AMI signal exhibits a notch at the spectral zero point. After passing through the PD, all positive and negative pulses are converted into positive values. By sampling at the correct timing, the original BPSK signal can be recovered through differential demodulation.

\begin{figure}[!h]
    \centering
    \includegraphics[width=1\textwidth]{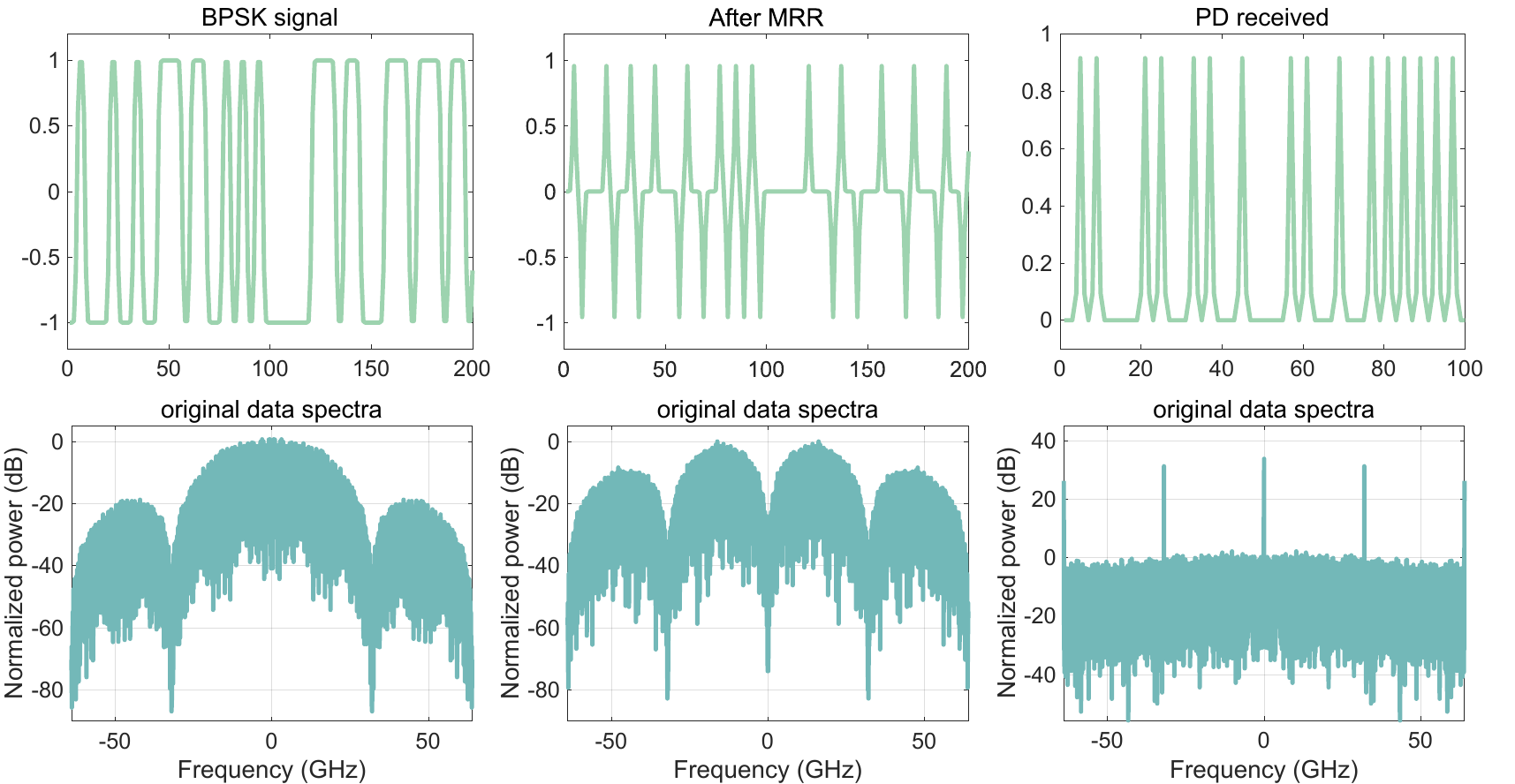}
    \caption{Simulation results of BPSK signals demodulation using MRR. }
    \label{fig:WDMbpsk}
\end{figure}

\noindent\textbf{Experiment setup}

More details are presented here for experiment setup. In our experimental setup, a signal to be processed is generated using an AWG (Keysight M8199A) with a sampling rate of 128 GSa/s. This signal is then modulated onto an optical carrier using a commercial lithium niobate (LiNbO$_3$) modulator(FTM7938EZ). Due to the grating coupler design employed for on-chip optical I/O ports, a polarization controller (PC) is utilized to adjust the polarization of the input optical signal, thereby maximizing the on-chip coupling efficiency. The output optical signal is first amplified by an erbium-doped fiber amplifier (Amonics DWDM AEDFA-C-DWDM-23-B-FA) and subsequently detected by an lnGaAs photodetector (PD) with a 70 GHz bandwidth. Finally, the waveform data is captured using a high-speed real-time oscilloscope (Keysight UXR0502A). For demodulation of 5x25Gbps WDM BPSK signals, a continuous-wave (CW) optical carrier generated by a laser is first fed into a LiNbO$_3$ modulator. A sinusoidal signal with a frequency of 40 GHz, produced by AWG, is used to drive the modulator, generating a multi-wavelength source comprising five distinct wavelengths. The optical spectrum of this frequency comb source is observed and recorded using an OSA (YOGOKAWA AQ6370D). Subsequently, the multi-wavelength source is amplified by an EDFA before being loaded into the photonic integrated circuit for further processing. Subsequently, the multi-wavelength source is amplified by an EDFA before being loaded into the photonic integrated circuit for further processing. Subsequently, the demodulated multi-wavelength signals are routed through a wavelength-selective switch (II-VI WS-01000A-C-S-1-AA-00) to achieve wavelength-division multiplexing. The waveform of each wavelength-specific demodulated signal is individually recorded and processed using digital signal processing (DSP) techniques to complete the demodulation process. Finally, the BER curves are measured to evaluate the system's performance under varying signal conditions.\\

\noindent\textbf{Real-Time Error Correction for Photonic GPAC}

Analog computing systems, including the photonic GPAC presented here, are inherently susceptible to environmental variations and component imperfections that lead to performance degradation. To ensure accurate and reliable computation, particularly for solving ODEs, the core processing unit, based on a reconfigurable micro-ring resonator (MRR), must be maintained in the critical coupling state, enabling accurate temporal differentiation. To achieve this, we have implemented an active, real-time error correction system based on a gradient descent algorithm, implemented on a field-programmable gate array (FPGA).

The primary source of error is the drift of the MRR's resonant frequency and operating point, mainly due to temperature variations, to which the thermal phase shifters are highly sensitive. Other contributing factors include laser power and wavelength fluctuations, amplifier gain drift, and photodetector responsivity changes.  Error correction relies on the integrated on-chip photodetector (PD) at the drop port of the core processing unit. The PD current is directly proportional to the optical power at the drop port, providing a sensitive, real-time indicator of the MRR's resonance condition and its coupling state. Deviations from the maximum PD current, corresponding to critical coupling, signal a departure from the desired operating point.

The core processing unit, crucial for performing the differentiation operation in the GPAC's ODE solver, consists of a double-ring MRR coupled via Mach-Zehnder interferometers (MZIs). The electric field at the drop port, \(E_d\), can be described by the following transfer function:
\begin{equation}
E_d = \frac{-1j \cdot a_0 \cdot e^{j(\phi_1 + \phi_2 + \alpha + \gamma + \beta)/2} \cdot \cos(\alpha/2) \cdot \cos(\gamma/2) \cdot \cos(\beta/2) \cdot E_{in}}{D}
\label{eq:transfer_function}
\end{equation}
where
\begin{align*}
D &= a_0^2 \cdot e^{j(\phi_1 + \phi_2)} \cdot e^{j(\alpha + \gamma + 2\beta)/2} \cdot \sin(\alpha/2) \cdot \sin(\gamma/2) \\
&\quad - a_0 \cdot e^{j\phi_1} \cdot e^{j(\alpha + \beta)/2} \cdot \sin(\alpha/2) \cdot \sin(\beta/2) - a_0 \cdot e^{j\phi_2} \cdot e^{j(\gamma + \beta)/2} \cdot \sin(\gamma/2) \cdot \sin(\beta/2) + 1.0
\end{align*}

where \(E_{in}\) is the input electric field. \(a_0 = \exp(-\sigma L / 2)\) is the round-trip transmission coefficient of a single MRR ring, with \(\sigma\) being the optical field transmission loss coefficient and \(L\) the single ring length. \(\phi_1\) and \(\phi_2\) are the phase shifts of light after one round trip in the two single rings, controlled by phase shifters PS4/PS5 and PS6/PS7, respectively. \(\alpha\), \(\beta\), and \(\gamma\) are the phase differences between the two arms of the MZIs, controlled by phase shifters PS1, PS2, and PS3, respectively.

This complex transfer function highlights the multiple degrees of freedom available for controlling the MRR's operating state. For the MRR to function as a first-order differentiator, two key conditions must be met: 1) The MRR must be in the critical coupling state.  2) The resonant wavelength must match the input laser wavelength.

To simplify the control problem, we adopt the following strategy.  PS2 is set to a fixed value, determining the free spectral range (FSR) of the core processor (either single-ring or double-ring configuration). PS6, which has a similar function to PS4 in controlling the MRR phase shift and resonant wavelength, is kept constant at 0V.  PS1, which controls the input coupling ratio, is also set to a fixed value determined by the specific computational task. For the MRR to operate as a first-order differentiator, the through-port transfer function must exhibit a $\pm \pi/2$ phase shift near resonance, approximated as '$j$' in the $j(\omega - \omega_0)$ term. This phase condition, derived from the full transfer function, dictates a specific relationship between the input and output coupling coefficients and is intrinsically linked to the round-trip phase accumulation within the ring. This leaves PS3, which controls the output coupling ratio, and PS4, which primarily affects the MRR phase shift and resonant wavelength, as the active control variables. Therefore, the combined optimization of PS3 and PS4 can meet both the critical coupling and resonant wavelength matching conditions of the MRR.

\begin{figure}[htbp]
  \centering
  \begin{subfigure}[b]{0.4\textwidth}
    \centering
    \includegraphics[width=\textwidth]{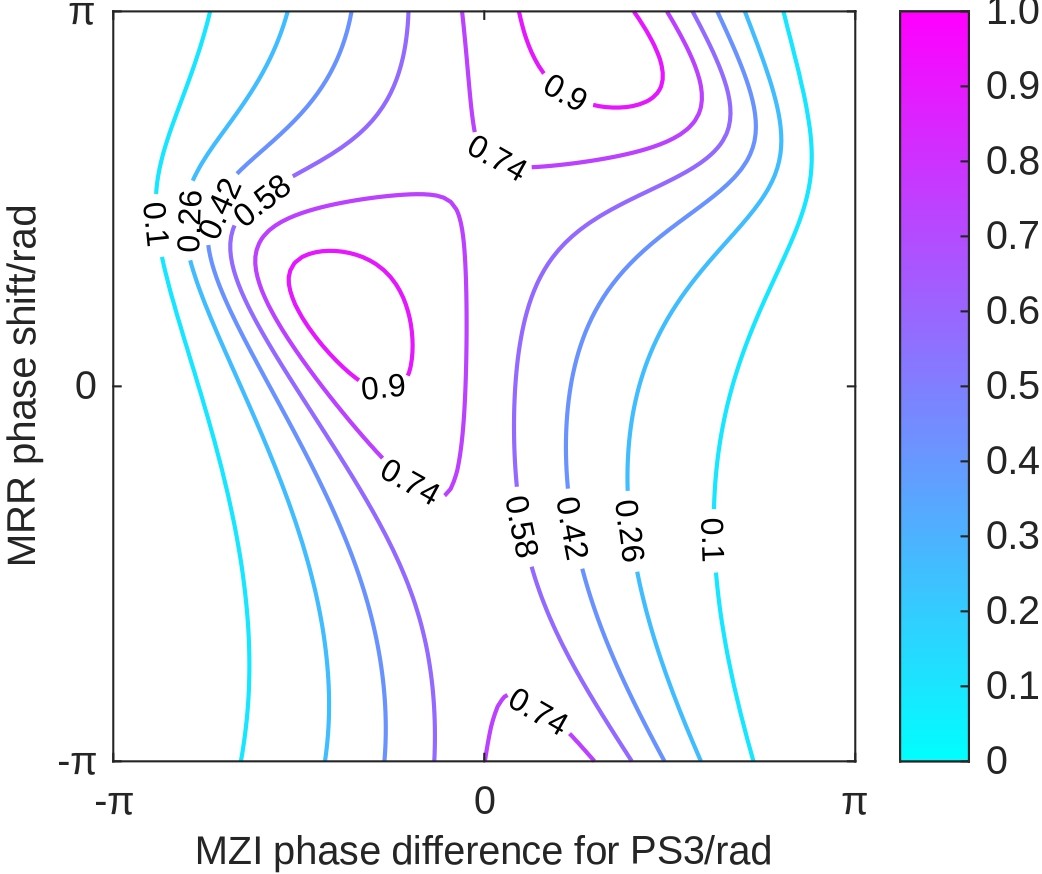}
    \caption{PS1 @ 0.3$\pi$}
    \label{fig:plot_0_3}
  \end{subfigure}
  \hspace{0.5cm} % Fixed horizontal space
  \begin{subfigure}[b]{0.4\textwidth}
    \centering
    \includegraphics[width=\textwidth]{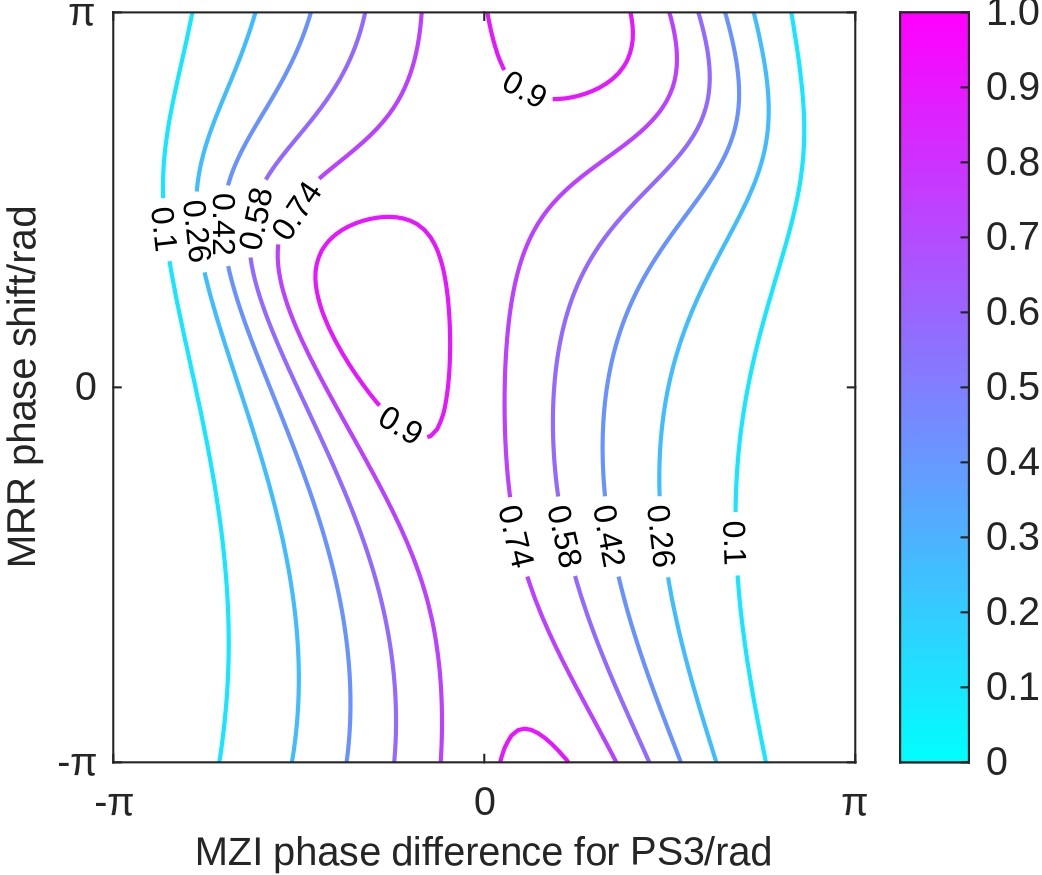}
    \caption{PS1 @ 0.2$\pi$}
    \label{fig:plot_0_2}
  \end{subfigure}

  \vspace{0.2cm}

  \begin{subfigure}[b]{0.4\textwidth}
    \centering
    \includegraphics[width=\textwidth]{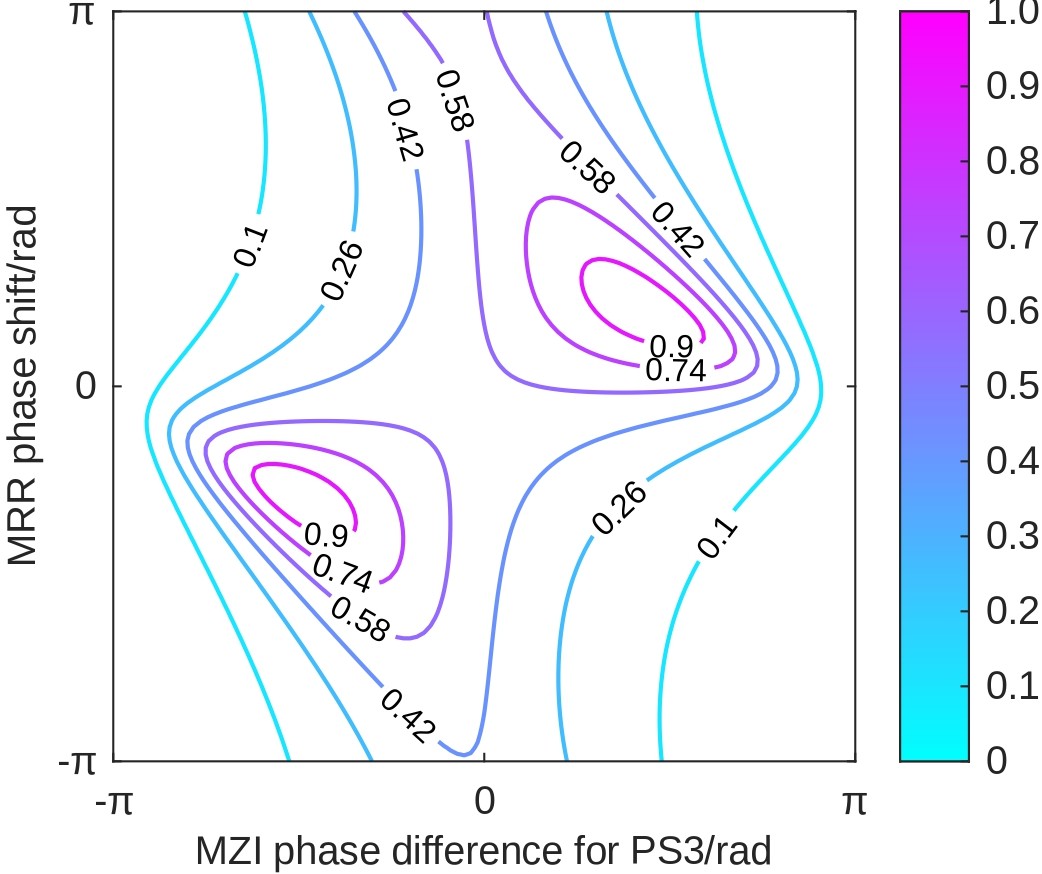}
    \caption{PS1 @ -0.5$\pi$}
    \label{fig:plot_-0_5}
  \end{subfigure}
  \hspace{0.5cm} % Fixed horizontal space
  \begin{subfigure}[b]{0.4\textwidth}
    \centering
    \includegraphics[width=\textwidth]{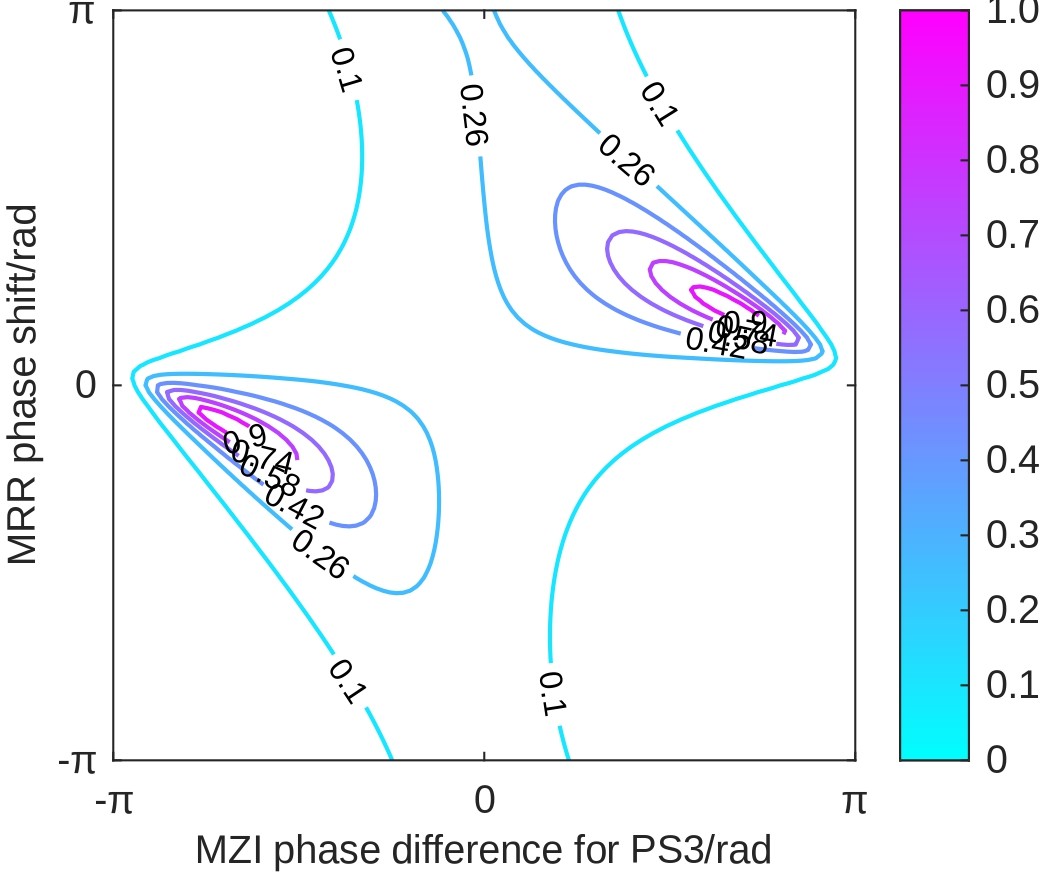}
    \caption{PS1 @ -0.7$\pi$}
    \label{fig:plot_-0_7}
  \end{subfigure}

  \caption{Normalized drop port power versus MZI phase difference (controlled by PS3 and PS4) for different fixed PS1 settings.}
  \label{fig:four_plots}
\end{figure}

Figure~\ref{fig:four_plots} illustrates the dependence of the normalized drop port power on the voltages applied to PS3 (Va) and PS4 (Vb) for four different fixed settings of PS1 (represented by phase differences of -0.7$\pi$, -0.5$\pi$, 0.3$\pi$, and 0.4$\pi$ between the MZI arms), with PS2 set to a 0.1$\pi$ phase difference, selecting the double-ring configuration. These figures demonstrate two crucial points.  First, the power landscape is smooth with respect to Va and Vb. This smoothness justifies the use of a gradient descent algorithm for optimization. Second, for each fixed PS1 setting, there exist periodic maximum power points, which can be reached by adjusting only PS3 and PS4. Therefore, by maximizing the PD current (proportional to the drop port power) using a gradient descent algorithm that adjusts Va and Vb, we can reliably bring the MRR to the critical coupling state and with the resonant wavelength matching the input laser wavelength, thus ensuring accurate first-order differentiation.

\begin{figure}[!h]
  \centering
  \includegraphics[width=0.7\textwidth]{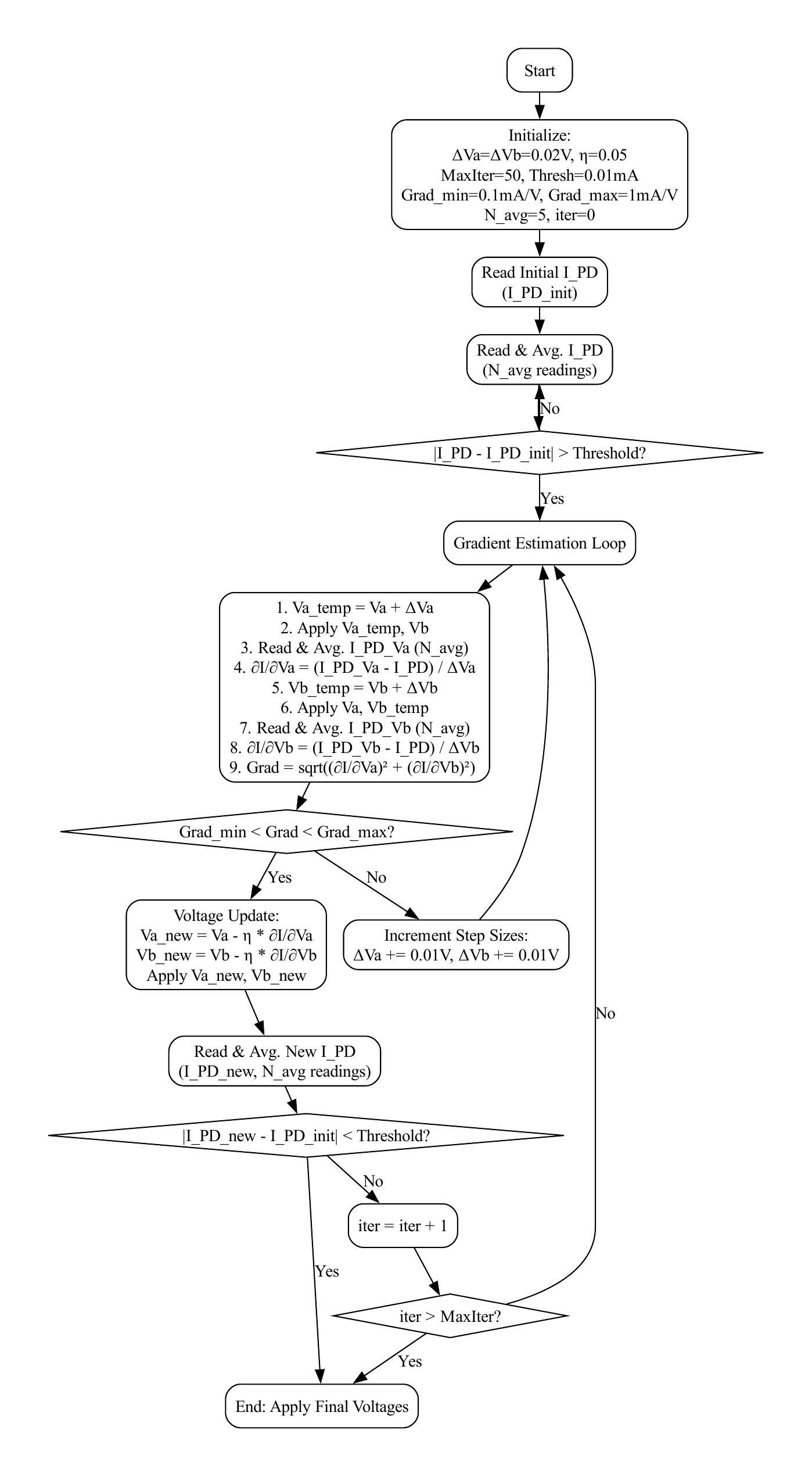} %
  \caption{Flowchart of the Real-Time Error Correction Algorithm Based on Gradient Descent.}
  \label{fig:gradient_descent_flowchart}
\end{figure}

The gradient descent algorithm iteratively adjusts Va and Vb according to the following update equations:

\begin{align}
V_{a\_new} &= V_a - \eta \cdot \frac{\partial I}{\partial V_a} \\
V_{b\_new} &= V_b - \eta \cdot \frac{\partial I}{\partial V_b}
\label{eq:gradient_descent}
\end{align}

where \(I\) is the PD current, and \( \eta \) is the learning rate. The partial derivatives are estimated numerically using finite differences:

\begin{align}
\frac{\partial I}{\partial V_a} &\approx \frac{I_{PD}(V_a + \Delta V_a, V_b) - I_{PD}(V_a, V_b)}{\Delta V_a} \\
\frac{\partial I}{\partial V_b} &\approx \frac{I_{PD}(V_a, V_b + \Delta V_b) - I_{PD}(V_a, V_b)}{\Delta V_b}
\label{eq:finite_diff}
\end{align}

where \(\Delta V_a\) and \(\Delta V_b\) are small voltage step sizes. An adaptive step size mechanism is employed, starting with \(\Delta V_a = \Delta V_b = 0.02\) V and increasing by 0.01 V increments if the gradient magnitude, \(Grad = \sqrt{(\partial I / \partial V_a)^2 + (\partial I / \partial V_b)^2}\), falls outside a predefined range (\(Grad_{min}\) to \(Grad_{max}\)). This ensures that the gradient estimate is not dominated by noise (step size too small) or averaging effects (step size too large). The algorithm is triggered when the measured PD current deviates from its initial maximum value (obtained after the initial configuration of PS1-PS6) by more than a threshold value. The algorithm then iterates, adjusting Va and Vb, until the PD current is within 0.01 mA of its initial maximum, or until a maximum number of iterations (50) is reached. To mitigate the effects of high-frequency noise, the PD current, I\_PD, used in the calculations, is the average of five consecutive readings from the Keithley 2450 source meter. Table 1 summarizes the key parameters of the algorithm. Figure \ref{fig:gradient_descent_flowchart} illustrates the flowchart of the real-time error correction algorithm for the photonic GPAC, which is based on gradient descent.

\begin{table}[!h]
\centering
\caption{Gradient Descent Algorithm Parameters}
\begin{tabular}{cccl}
\botrule
Parameter & Symbol & Value & Justification \\
\midrule
Initial Voltage Step Size (Va) & \(\Delta V_{a\_initial}\) & 0.02 V & Balance between noise and accuracy \\
\midrule
Initial Voltage Step Size (Vb) & \(\Delta V_{b\_initial}\) & 0.02 V & Balance between noise and accuracy \\
\midrule
Voltage Step Size Increment & \(\Delta V_{increment}\) & 0.01 V & Adaptive step size adjustment \\
\midrule
Learning Rate & \(\eta\) & 0.05 & Balance between speed and stability \\
\midrule
Maximum Iterations & \(Max\_Iterations\) & 50 & Prevent infinite loops \\
\midrule
PD Deviation Threshold & \(Threshold\) & 0.01 mA & Target accuracy \\
\midrule
Minimum Gradient Magnitude & \(Grad_{min}\) & 0.1 mA/V & Avoid noise-dominated gradients \\
\midrule
Maximum Gradient Magnitude & \(Grad_{max}\) & 1 mA/V & Avoid averaging effects \\
\midrule
Number of Averaged Samples & \(N_{avg}\) & 5 & Reduce high-frequency noise \\
\midrule
Sampling Rate &  & 50 Hz &  Nyquist frequency consideration \\
\midrule
Target Update Rate &  & \(\geq\) 5 Hz & Ensure the real time and rapid \\
\botrule
\end{tabular}
\label{tab:parameters}
\end{table}

The algorithm is implemented on an ALINX Xilinx Artix-7 CZ3452 FPGA, operating at a clock frequency of 200 MHz. A block diagram of the implementation is shown in Fig \ref{fig:gradient_descent_flowchart}. The key modules include a Keithley 2450 interface for reading the PD current via Ethernet, a 128-channel voltage source control module utilizing a USB interface, a gradient calculation module (implementing Equations \ref{eq:gradient_descent} and \ref{eq:finite_diff} and averaging), a voltage update module, and a control logic module that manages the threshold checks and iteration count. The system utilizes floating-point arithmetic for ease of development, the gradient calculation and voltage update modules share a single floating-point divider and multiplier, as these operations are not performed concurrently.

The data flow begins with the FPGA receiving the averaged PD current from the Keithley 2450 via Ethernet. An integrated Ethernet MAC and custom protocol handler manage the communication, acquiring the averaged PD current data for the gradient calculation module.

The gradient calculation module is responsible for computing the partial derivatives  \(\frac{\partial I}{\partial V_a}\) and \(\frac{\partial I}{\partial V_b}\) using the finite difference method. To mitigate the effects of quantization noise and improve the accuracy of the gradient estimation, a numerically stable implementation of the finite difference calculation is employed. This involves careful consideration of the order of operations and the use of appropriate scaling factors to prevent overflow or underflow in the floating-point calculations. The adaptive step size mechanism, controlled by the  \(Grad_{min}\) and \(Grad_{max}\) thresholds, is implemented using a state machine within the control logic module.

The voltage update module implements the gradient descent update equations (Equations \ref{eq:gradient_descent}). The learning rate \( \eta \) is implemented as a fixed scaling factor applied to the calculated gradients. To ensure stability and prevent oscillations, the updated voltage values, \(V_{a\_new}\) and \(V_{b\_new}\), are checked against predefined upper and lower bounds before being sent to the voltage source. This prevents the application of voltages outside the operational range of the phase shifters.

The 128-channel voltage source is controlled via a USB interface. The FPGA implements a USB device controller and custom protocol handler to manage communication and apply voltage updates to the selected channels. A double-buffering scheme ensures glitch-free transitions between voltage updates.

The control logic module orchestrates the entire error correction process. It implements the state machine that controls the sequence of operations, including reading the PD current, calculating the gradient, updating the voltages, checking for convergence, and managing the iteration count. The control logic also handles the adaptive step size mechanism and the threshold checks for both the PD current deviation and the gradient magnitude.

The control loop achieves an update rate of approximately 8 Hz, well within the target of at least 5 Hz, ensuring rapid compensation for drift. This update rate is limited primarily by the communication and settling time of the 128-channel voltage source via USB (approximately 300 ms). The computation time within the FPGA itself is negligible in comparison.

To evaluate the performance of the real-time error correction system, experiments were conducted using a setup comprising the fabricated photonic GPAC chip, a Keithley 2450 source meter for high-precision PD current measurement, a 128-channel voltage source for controlling the phase shifters, and an ALINX Xilinx Artix-7 CZ3452 FPGA implementing the error correction algorithm. The GPAC was configured to perform a first-order differentiation of a Gaussian input pulse with a FWHM of 200 ps.

The effectiveness of the real-time error correction system was experimentally verified on the fabricated photonic GPAC chip, configured for first-order differentiation.  As previously discussed and illustrated in Figure \ref{fig:control}(b), the error correction algorithm successfully maintained the PD current within a narrow tolerance (RMS deviation of 0.005 mA) over a 30-minute period, representing an 89\% improvement in stability compared to the uncorrected system (RMS deviation of 0.046 mA).  This stabilization of the PD current, and thus the MRR operating point, ensured the accurate generation of the first-order derivative waveform (Figure \ref{fig:control}(c)), demonstrating the system's ability to mitigate the effects of environmental drift and maintain reliable analog computation.\\

\noindent\textbf{Energy consumption analysis of the GPAC}

Here, we evaluate and compare the energy efficiency of the chip's processing capabilities. The core processing units on the chip primarily rely on thermally tuned phase shifters (PS) to configure their operational states. Silicon-based photonic chips offer a distinct advantage over other platforms, such as thin-film lithium niobate (TFLN), in terms of relatively lower power consumption. As evidenced by the baseline test results in Fig~\ref{fig:chip setup}(c), the titanium nitride (TiN) heaters in the silicon photonic chip exhibit an efficiency of approximately 2.2 mW/$pi$, whereas thermally tuned phase shifters in thin-film LiNbO$_3$ chips typically require hundreds of mW/$pi$. Taking the demodulation of BPSK signal as an example, when performing single-channel and single-wavelength demodulation, only one core processing unit on the chip is in the working state. The MZI optical switches within the front-and-rear Optical matrices are configured in the Cross state to minimize inter-channel crosstalk. Under these conditions, the signal processing speed reaches 32 giga-operations per second (GOPS), while the total power consumption of the on-chip thermally tuned phase shifters is approximately 12 mW. Consequently, the chip achieves an energy efficiency of approximately 2.67 TOPS/W. 

Moreover, when the GPAC photonic chip performs demodulation of multi-wavelength BPSK signals through frequency-division multiplexing, it does not increase the power consumption of the chip. The reconfigurable MRR still operates in the state with a FSR of 40 GHz, yet the processing speed is multiplied. For instance, in the experiment validating the demodulation of 5×25 Gbps WDM BPSK signals, the processing speed of the chip in this experiment is 125 GOPS. Thus, the energy efficiency of the chip is approximately 10.42 TOPS/W. Theoretically, if multi-wavelength BPSK signals span the entire C-band, 109 signals will be demodulated simultaneously. During the demodulation process, the chip can achieve a processing speed of 2.725 TOPS, with an energy efficiency of approximately 227 TOPS/W.

\section*{Declarations}

% Some journals require declarations to be submitted in a standardised format. Please check the Instructions for Authors of the journal to which you are submitting to see if you need to complete this section. If yes, your manuscript must contain the following sections under the heading `Declarations':

\begin{itemize}
\item Funding

This work is supported by the Pengcheng Laboratory Major Key Project, National Talent Program, and National Natural Science Foundation of China (62220106002, 62125103, 62171059, 62401082).

\item Conflict of interest/Competing interests

The authors declare no competing interests.

\item Ethics approval and consent to participate
\item Consent for publication
\item Data availability 
\item Materials availability
\item Code availability 
\item Author contribution

Z.Z.C. conceived the concept of reliable general-purpose analog computing and created the concrete on-chip processor unit architecture. T.Z. designed the basic silicon photonic element and completed the circuit layout. T.Z. and S.C.Z. performed the basic measurements of the chip. S.C.Z., X.W.X. and S.G.H. led the design for FPGA based real-time error-correction algorithm and implemented the codes. B.W.Z. designed the system experimental plan and developed the DSP algorithm for demonstrated applications, T.Z. and B.W.Z. carried out the high-speed measurements with the help of Y.Z.P. and analyzed the data. J.Y. completed the PCB assembly. T.Z., B.W.Z. and K.R.L. prepared the manuscript with contributions from all authors. Z.Z.C. led the whole research, Z.Z.C., X.C.X and S.H.Y. supervised the project. T.Z., B.W.Z. and S.C.Z. contributed equally to this work. 
\end{itemize}

%\noindent
%If any of the sections are not relevant to your manuscript, please include the heading and write `Not applicable' for that section. 

%\bibliography{bibliography}% common bib file

\end{document}